\documentclass[11pt]{iopart}
\usepackage{iopams} % comment out if AMS fonts not required
\usepackage[utf8]{inputenc}
\usepackage[T1]{fontenc}
\usepackage{subcaption}
\usepackage{longtable,booktabs,array}
\usepackage[backend=biber, sorting=none, style=numeric-comp]{biblatex}
\usepackage[switch]{lineno}
\usepackage{graphicx}
\usepackage{hyperref}

\hypersetup{colorlinks=true, breaklinks=true, %pagebackref=true,
  urlcolor=blue, linkcolor=blue,anchorcolor=blue,citecolor=blue,
  hypertexnames=true, final=true, 
  pdfpagemode = UseNone, %FullScreen, %UseThumbs, %UseOutlines,
  pdfauthor = {},
  pdftitle = {},   
  pdfsubject = {},
  pdfkeywords = {}
}

\graphicspath{{imglocal} {./}}

\begin{document}

\title[Editorial]{Introducing ``Neuromorphic Computing and Engineering''}

\author{Giacomo Indiveri$^1$}

\address{$^1$ Institute of Neuroinformatics, University of Zurich and ETH Zurich, Zurich, Switzerland}
\ead{giacomo@ini.uzh.ch}

\begin{abstract}
  The standard nature of computing is currently being challenged by a range of problems that start to hinder technological progress. One of the strategies being proposed to address some of these problems is to develop novel brain-inspired processing methods and technologies, and apply them to a wide range of application scenarios. This is an extremely challenging endeavor that requires researchers in multiple disciplines to combine their efforts and co-design at the same time the processing methods, the supporting computing architectures, and their underlying technologies. The journal ``Neuromorphic Computing and Engineering'' (NCE) has been launched to support this new community in this effort and provide a forum and repository for presenting and discussing its latest advances. Through close collaboration with our colleagues on the editorial team, the scope and characteristics of NCE have been designed to ensure it serves a growing transdisciplinary and dynamic community across academia and industry.
\end{abstract}

%\keywords{winner-take-all, device mismatch, population coding, spiking neural network, recurrent neural network}
\submitto{Neuromorphic Computing and Engineering}
\maketitle

% \ioptwocol % Use with \documentclass[10pt]{iopart} for final version preview

\section{Introduction}
\label{sec:introduction}

As Editor-in-Chief I'm pleased to announce the publication of the first content~\cite{Musisi-Nkambwe_etal21,Leroux_etal21,Yan_etal21,Jaeger21} in
\emph{Neuromorphic Computing and Engineering} (NCE).
This editorial aims at motivating the need for creating such a journal by describing our view on some of the current challenges for Information and Communication Technologies (ICT) and the role that NCE can play to address them.

\begin{figure}
  \begin{subfigure}{0.65\textwidth}
    \centering
    \includegraphics[width=0.75\textwidth]{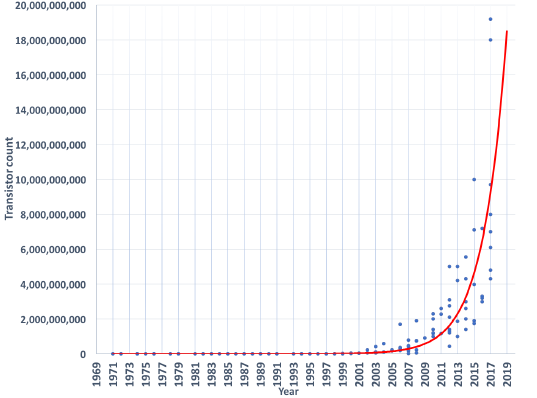}
    \subcaption{} 
    \label{fig:mooreslaw}
  \end{subfigure}
  \begin{subfigure}{0.35\textwidth}
    \centering
    \includegraphics[width=0.75\textwidth]{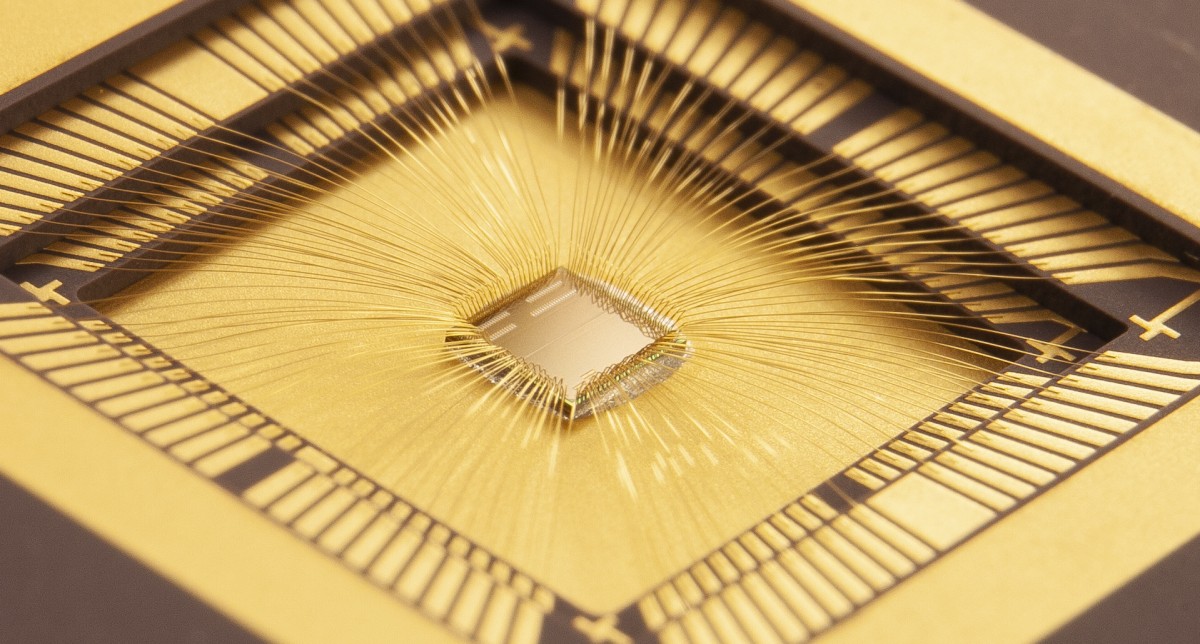}
    \subcaption{} 
    \label{fig:dynapse}
  \end{subfigure}
  \caption{Technological progress. In computers, (\subref{fig:mooreslaw}) Number of transistors integrated on a single CMOS chip over the years (source  \href{https://bjc.edc.org/March2019/bjc-r/course/bjc4nyc.html}{``The Beauty and Joy of Computing—BJC''}; (\subref{fig:dynapse}) Example of a recent neuromorphic CMOS VLSI chip comprising thousands of silicon neurons and dynamic synapses (source \href{http://www.ini.uzh.ch}{Institute of Neuroinformatics, University of Zurich and ETH Zurich}).}
  \label{fig:progress}
\end{figure}

\paragraph{} While Moore's law has been fueling technological progress for decades, it is only in recent years that we could start to reap most of this progress' benefits (see Fig.~\ref{fig:mooreslaw}): ICT are becoming pervasive and are affecting every facet of our daily lives.
At a global scale, ICT is affecting almost all aspects of our society, ranging from global communication to education and health, from finance to automation, transportation, and climate change.
With the digitalization of our society, computing technologies and electronic devices are producing increasing amounts of electronic data every year and people are getting access to larger and larger amounts of personalized information.

To cope with the demands that are emerging with this technological revolution, and to exploit the opportunities created with the availability of this data, 
novel data-science methods, machine-learning techniques, and Artificial Intelligence (AI) algorithms have emerged.
AI algorithms typically employ neural networks and deep learning techniques to solve pattern recognition tasks and have been shown to be extremely successful in extracting information from large amounts of data~\cite{LeCun_etal15,Schmidhuber15}.
However,
% despite their tremendous success, AI and machine learning algorithms face some serious challenges that severely limit their ability to improve and to scale: 
% the fundamental reasons for which these algorithms can achieve such impressive results are still poorly understood~\cite{Sejnowski20}, they can be trained to solve only a very limited set of problems at a time (``narrow AI''), and
their training methods require massive amounts of data and computing resources, which in turn requires an amount of energy that is not sustainable with respect to both the global electricity supply and the computation's Carbon footprint. 
For example, it has been estimated that the time required to train a recent state-of-the-art AI neural network, such as GPT-3~\cite{Floridi_Chiriatti20}, would take more than 27 years worth of processing time on a single computer, and that these computations would generate over 78,000 pounds of $CO_2$ emissions in total, which is more than what the average American adult produces in two years.\footnote{from the Forbes article \href{https://www.forbes.com/sites/robtoews/2020/06/17/deep-learnings-climate-change-problem/}{``Deep Learning’s Carbon Emissions Problem''}.}
The reason for these exorbitant costs are due to the fact that current computing technologies, based on the classical von Neumann architectures, are not well matched to the parallel processing nature of neural networks~\cite{Indiveri_Liu15}.
% They are based on the classical von Neumann architectures, built using separate memory and processing units that transfer data across a shared serial bus as quickly as possible.
%From the technology point of view, conventional \emph{von Neumann} computing architectures used to run these algorithms are not ideally suited for the parallel nature of neural networks. 
% Their standard mode of operation, based on the use of separate memory and processing units, results in a communication bottleneck that produces exceedingly large power consumption.
On the other hand, biological brains clearly outperform AI systems in terms of the amount of power consumption requirements, the number of training data samples, and their ability to adapt to novel and unexpected conditions.
This is particularly true for those tasks that are still very difficult for computers and AI algorithms, but are done effortlessly by humans and animals, such as online learning, with small numbers of examples, interaction with the environment, or sensing and motor control.

%Although deep learning is inspired by the architecture of the cerebral cortex, and despite the tremendous progress neuroscience made in understanding the principles of computation used by the brain, there is still a huge gap in relating both the physical computing substrate and the organizing principles of natural intelligence to AI computing systems.
A promising approach toward the development of novel computational paradigms and ICT systems that can interact intelligently with the environment, that could bridge the gap between natural intelligence and artificial intelligence, and that could solve many of the open challenges facing the future of computing is the one pursued by the ``neuromorphic computing and engineering'' field. 

We are now at a very exciting time in which the convergence of the end or Moore's law, with the renewed interest in neural networks, and the needs for low power and sustainable ``Green-AI'' are all pointing toward the huge potential that can come from research and development in neuromorphic computing and engineering.

\section{NCE thematic areas}
\label{sec:areas}

The term \emph{neuromorphic} was originally coined by Carver Mead in the late 1980s at CALTECH, to describe Very Large Scale Integration (VLSI) systems containing electronic circuits that  mimic neuro-biological architectures present in the nervous system~\cite{Mead20,Mead90}.
However, throughout the years, the term has ``morphed'' its original meaning and started to be used to describe a broader set of concepts, and approaches. Next to the original ``neuromorphic engineering'' meaning, the term started to be used to describe ``neuromorphic computing'' systems that comprise also pure digital circuits or conventional processors, for simulating spiking neural networks and neural models of computation.
In parallel, the same term started to be used to refer to systems comprising nano-scale ``memristive'' devices, developed within the field of emerging memory technologies.
Today, the term is also being used to refer to algorithmic and machine learning approaches that simulate biologically plausible and hardware friendly spiking neural networks and learning mechanisms. 

\subsection*{Neural circuits and systems integrated in CMOS technology}

The main goal of the original neuromorphic engineering approach was to directly \emph{emulate} the physics of computation of biological neural networks, via the use of transistors operated in their ``weak-inversion'' or ``subthreshold'' regime~\cite{Mead89,Douglas_etal95a,Liu_etal02a}.
This approach was aimed at building artificial neurons, synapses, networks, and sensory systems using the same organizing principles used by the nervous system in animal brains.
This effort had the dual objective of both understanding neural computation by building physical emulations of real neural circuits, and developing compact and low power sensory processing and computing architectures radically different from the standard digital computers of the time.
Given the high-risk and basic research aspect of this approach, there is only a small number of academic groups that are still pursuing it today. This community is mostly focusing on the development of small-scale prototype chips to explore different aspects of neural computation, ranging from sensory systems~\cite{Kramer02a,Lichtsteiner_etal06a,Wen_Boahen09,Liu_Delbruck10,Liu_etal14a} to reconfigurable networks with biologically plausible neural dynamics~\cite{Benjamin_etal14,Qiao_etal15,Moradi_etal18,Park_etal16}, to spike-based learning and plasticity circuits~\cite{Mahvash_Parker13,Banerjee_etal15,Huayaney_etal16,Detorakis_etal18,Qiao_etal17}.

%\subsection*{The large-scale system simulation approach}

In more recent times the term neuromorphic was adopted also to describe mixed-signal and pure digital VLSI systems which implement computing platforms that could be used to \emph{simulate} models of spiking neural networks.
This effort was mainly driven by the possibility to exploit progress in computing and integrated circuit technology for building large-scale dedicated brain-inspired computing systems.
For example, the EU Human Brain Project funded the development of wafer-scale integrated systems designed to faithfully reproduce simulations of neuroscience modeling studies comprising large numbers of neurons at accelerated speeds~\cite{Schemmel_etal10}.
Similarly, the SpiNNaker system, also developed with the support of the Human Brain Project, is a multi-core computer designed with the goal of simulating very large numbers of spiking neurons in real time~\cite{Furber_etal14}. At the current state of development the SpiNNaker Machine, built by stacking together 600 Printed Circuit Boards, each comprising 48 SpiNNaker processors, supports the simulation of hundreds of millions of neurons.
% One early example, that used the CALTECH approach of using analog subthreshold circuits to emulate neural dynamics, was the ``NeuroGrid'' system~\cite{Benjamin_etal14}.
%This neuromorphic setup combined into a single printed circuit board 16 VLSI chips, each comprising 256$\times$256 (analog) silicon neurons, communicating among each other, and across chips, using the asynchronous digital ``Address-Event Representation'' (AER).
%This was the first example of a complex neuromorphic system reaching the size of one million neurons.
An alternative strategy to scale up the size of simulated spiking neural networks is the one proposed by IBM, which in 2014 presented the ``TrueNorth'' neuromorphic systems that integrated on the same chip 4096 cores, each comprising pure digital asynchronous circuits able to simulate 256 neurons with 256$\times$256 synaptic connections~\cite{Merolla_etal14a}.
This was a significant breakthrough in the field, as it demonstrated how advanced technology nodes, such as the one used of the Samsung 28\,nm bulk CMOS process, could support the integration of very large numbers of silicon neurons, while keeping the overall power consumption extremely low (e.g., with an average consumption of 70\,mW total power while running a typical recurrent network at biological real-time, four orders of magnitude lower than a conventional computer running the same network).
The more recent Loihi chip built by Intel~\cite{Davies_etal18} has some common design choices with the IBM TrueNorth chip, in that it uses pure digital asynchronous circuits to simulate neurons and synapses.
Taking advantage of the progress in integration technology, this chip was fabricated using the Intel 14\,nm FinFET process.
Rather than focusing on large-sale, the Loihi designers chose to focus on more complex neural and synapse features, including spike-based learning mechanisms.
So the Loihi chip integrates ``only'' 128 cores, and supports the simulation of networks containing up to 130'000 spiking neurons.
Similar to the SpiNNaker and the TrueNorth chips, Loihi is being used as a research platform to support the development of spike-based processing architectures that can be applied to the solution of practical problems.
Indeed, researchers and students are encouraged to develop novel spike-based computing solutions using Loihi, via the support of the Intel Neuromorphic Research Community (INRC) program, which has been very successful in promoting the growth of the community and producing a range of promising results~\cite{Davies_etal21}.

\subsection*{Memristive devices and emerging memory technologies}

The material science and device physics community has been carrying out research on new materials and technologies for memory and long-term storage applications for many years. Recently however, this community started using the term ``neuromorphic'' to refer to new devices and systems that, on one hand exhibit different types of behaviors that can be linked to those of biological synapses, and on the other represent important building blocks for the development of large-scale AI computing systems~\cite{Di-Ventra_etal09}.
Indeed, shortly after the proposal of using these new nano-scale devices as ``memristors''~\cite{Strukov_etal08}, this community quickly embraced the idea that these  devices could be used to implement synapses in artificial neural networks, and to store locally their synaptic weight~\cite{Jo_etal10}.
These devices and technologies hold a great promise of enabling ``in-memory computing'' in neural networks, and of supporting, through their physics, complex non-linear features that could be exploited to emulate many interesting properties of biological synapses.
Within this context, a wide range of research efforts are being pursued, including the development of different types of non-volatile and volatile memristive devices, and the design of spike- or pulse-based control schemes for inducing biologically plausible learning behaviors in memristive cross-bar arrays~\cite{Zamarreno-Ramos_etal11,Saighi_etal15}.
As there is not yet a single solution to the problem of finding the optimal artificial synapse, a broad range of materials, devices, and techniques are still an active area of investigation~\cite{Waser_Aono07,Yang_etal08,Chanthbouala_etal12,Ambrogio_etal16,Salinga_etal18,De-Burgt_etal18,Sebastian_etal19}.

\subsection*{Algorithms and computational models of spike-based learning and inference}

An important aspect of the neuromorphic computing and engineering domain is the one of hardware-software \emph{co-design}. It is no surprise then that the term neuromorphic is also being used to describe research in computational models and algorithms that can be readily mapped onto memristive, CMOS, or hybrid memristive-CMOS neuromorphic architectures~\cite{Chicca_Indiveri20}.
Most efforts in this domain focus either on exploring spike-based learning methods that approximate the backpropagation algorithm~\cite{Neftci_etal19,Bellec_etal20}, or on identifying local stochastic and complex non-linear plasticity mechanisms that can be reproduced by memristive devices or CMOS learning circuits~\cite{Payvand_etal20,Payvand_etal19,Brivio_etal19,Chicca_Indiveri20,Laborieux_etal21}. Very promising results are also being obtained by combining the latest advancements in AI and machine learning algorithms with brain-inspired and computational neuroscience modeling efforts~\cite{Berdan_etal16,Diederich_etal18,Serrano-Gotarredona_etal13,Lillicrap_Santoro19}. These investigations provide useful specifications for the design of new non-volatile and volatile memristive devices and for the design of spiking neural network chips that can use the principles of computation derived from the brain to carry out low-power and robust computation using local learning rules and unreliable and low-precision components, such as the ones present in animal brains.

\section{Open challenges}
\label{sec:challenges}

\begin{figure}
  \centering
  \includegraphics[width=0.35\textwidth]{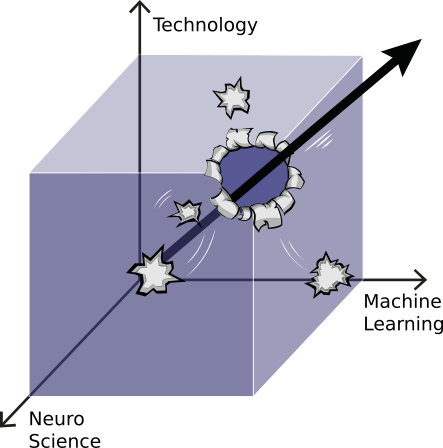}
  \caption{Thinking outside the box: NCE will promote and inspire new, unconventional, and innovative ideas in neuromorphic computing and engineering by supporting cross-fertilization and promoting the convergence of the disciplines and thematic areas that characterize this emerging research field.}
  \label{fig:ncebox}
\end{figure}

The most important challenge that NCE faces is that of supporting a diverse and interdisciplinary community of researchers working on different aspects of neuromorphic computing and engineering. This journal will be instrumental in aligning the goals and objectives of this community, and in promoting its growth (see Fig.~\ref{fig:ncebox}). The success of NCE strongly depends on how much the experts in the different thematic areas will be willing to broaden their horizon, learn the ``language'' of the experts of other thematic areas, and create synergies with them.
This is indeed already happening to some degree: material science and device experts are collaborating with circuit designers to both extend current CMOS technologies and integrate nano-scale devices in newly developed ones; computer science and machine learning experts are collaborating with neuroscientists to develop brain-inspired processing theories that are compatible with the neuromorphic technologies being developed; electronic engineers, physicists and memory device experts are working closely with computational neuroscientists to implement devices and circuits that can emulate the biophysics of synapses and neurons onto the electronic circuits and systems being developed.
By defining a common vision and by bridging the traditional boundaries between standard subjects, the NCE journal will enable the development and dissemination of breakthroughs in neuromorphic computing and engineering that will have disruptive effects, and which could revolutionize the nature of computation (see ``thinking outside the box'' Fig.~\ref{fig:ncebox}).

\section*{Conclusions and outlook}
\label{sec:outlook}

Neuromorphic computing and engineering is becoming an extremely important and timely research area. Basic research in brain-inspired neuromorphic electronic circuits has been carried out already for a significant number of years. In parallel tremendous progress has been made in computational neuroscience, nanotechnologies, and  machine learning. This is the best time to combine the know-how gained so far in these disciplines to obtain breakthroughs that can potentially solve many of the problems that ICT are starting to face.
The NCE journal has been established to provide a new open access medium explicitly designed to support this convergence of efforts, and to bring together all researchers engaged in these areas in a way that unites the community and defines the field of ``neuromorphic computing and engineering'' for years to come.

\printbibliography

\end{document}